# mqr-tree: A 2-dimensional Spatial Access Method

Marc Moreau and Wendy Osborn

**Abstract**—In this paper, we propose the mqr-tree, a two-dimensional spatial access method that organizes spatial objects in a two-dimensional node and based on their spatial relationships. Previously proposed spatial access methods that attempt to maintain spatial relationships between objects in their structures are limited in their incorporation of existing one-dimensional spatial access methods, or have lower space utilization in its nodes, and higher tree height, overcoverage and overlap than is necessary. The mqr-tree utilizes a node organization, set of spatial relationship rules and insertion strategy in order to gain significant improvements in overlap and overcoverage. In addition, other desirable properties are identified as a result of the chosen node organization and insertion strategies. In particular, zero overlap is achieved when the mqr-tree is used to index point data. A comparison of the mqr-tree insertion strategy versus the R-tree shows significant improvements in overlap and overcoverage, with comparable space utilization. In addition, a comparison of region searching shows that the mqr-tree achieves a lower number of disk accesses in many cases.

◆

## 1 INTRODUCTION

MANY applications existing today that store and manipulate spatial data. A spatial database [22] contains a large collection of objects that are located in multi-dimensional space. For example, the Geological Survey of Canada maintains a repository of spatial data for many geosciences applications [14], while the Protein Data Bank [6] contains many three-dimensional protein structures.

An important issue in spatial data management is to efficiently retrieve objects based on their location by using spatial access methods. An approximation method is a spatial access method that maintains a hierarchy of approximations of both objects and the space occupied by subsets of objects. Approximations are usually represented using a minimum bounding rectangle. Many approximation strategies have been proposed in the literature. However, most proposed strategies do not preserve all spatial relationships between objects because the data, which is represented in n-dimensional space, is forced into a 1-dimensional ordering [3], [5], [9], [10], [11], [15], [20]. This leads to inefficient searching, both within a node and the structure as a whole, because the only option is a linear search of a node in its entirety.

A few strategies have been proposed that attempt to preserve spatial relationships between objects [1], [12], [16], [18]. However, they are limited in that they still incorporate existing 1-dimensional spatial access methods, bulk-load their objects only, or result in a very high tree height and low average space utilization. These limitations can lead to high coverage, overcoverage and overlap of minimum bounding rectangles.

Therefore, we propose the mqr-tree, which improves upon existing spatial access methods by proposing a modified node organization and set of spatial relationship rules, and new insertion and search algorithms. We evaluate the mqr-tree against a benchmark strategy, the R-tree [9]. We show that the mqr-tree achieves significant improvements in overlap and overcoverage over the R-tree. Also, we show that the mqr-tree can achieve a significantly lower number of disk accesses when performing a region search. With these improvements accomplished, different searching strategies can be explored that can perform a partial search of nodes.

This paper proceeds as follows. Section 2 presents related work in the area of spatial access methods and their limitations. Section 3 presents the mqr-tree, in particular its node organization, insertion strategy and region search strategy. Section 4 presents special properties of the mqr-tree. Section 5 presents the results of our experimental evaluation versus the R-tree. Finally, the paper concludes and gives research directions in Section 6.

## 2 RELATED WORK

Many approaches for indexing objects based on location are proposed in the literature (see [8], [19], [22] for surveys). These approaches are classified into three categories [8]: main memory methods, point access methods and spatial access methods (spatial access methods). Many important strategies are proposed in all categories. We focus on spatial access methods, since our work resides in this category.

Spatial access methods provide uniform access to both point and object data. Also, they remain height-balanced in the presence of a dynamic object set. Many spatial ac-

- M. Moreau is with the City of Calgary, Calgary, Alberta, Canada, This work was undertaken while author was a student at the University of Lethbridge.
- W. Osborn is with the Department of Mathematics and Computer Science, University of Lethbridge, Lethbridge, Alberta, Canada, T1K 3M4.
- Paper is extended version of [13] and [14], and contains modified material from those papers

cess methods are proposed in the literature [1], [2], [3], [5], [9], [10], [11], [12], [15], [16], [18], [20]. They can be classified [8] into approximation, clipping, and mapping methods.

Approximation methods store a hierarchy of approximations of both objects and the space occupied by subsets of objects. Since the space is not partitioned, approximations can overlap. Many approximation methods are proposed, including the R-tree [9], the R⁺-tree [3], the X-tree [5], the R⁺Q-tree [8], the PR-tree [1], the 2DR-tree [16], [17], the VoR-tree [21], the DR-tree [12] and the MSI approach [2]. Clipping methods, such as the R⁺-tree [20], partition an object into parts so that overlap is avoided. Mapping methods map objects in n-dimensional space into a one-dimensional order. The objects are then stored and retrieved using an access method such as a B⁺-tree [7]. Approaches that use mapping include Z-ordering [15], the Hilbert R-tree [10], and the Filter tree [11].

A limitation to most hierarchical spatial access methods is their one-dimensional structure. However, no n-dimensional to one-dimensional mapping of spatial data exists that preserves all spatial relationships between objects [8]. This forces objects in n-dimensional space to be in a one-dimensional ordering, which results in the loss of spatial relationships. This leads to inefficient searching, both within a node and the structure as a whole, because the only option is a linear search of a node in its entirety. Mapping methods do provide a one-dimensional ordering of objects, but they cannot maintain all spatial relationships.

A few proposed strategies that attempt to overcome this one-dimensional limitation are the R*-Q-tree, the 2DR-tree, the DR-tree and the MSI approach. In the R*-Q-tree, the space that contains objects is partitioned into four quadrants, and a standard R*-tree is constructed for each. Although the root level takes the spatial relationships of objects into account, the existence of the R*-trees still implies that one-dimensional mapping of objects in n-dimensional space still takes place. The MSI approach extends the R*Q-tree approach by recursively partitioning the quadrants until the number of objects in each region falls below a specified threshold, before constructing a traditional index for each region, such as an R-tree. The regions are stored in a metaTree to facilitate access to the required R-trees. Along with the R*-Q-tree, the MSI approach inherits the one-dimensional property of object organization, since the R-tree is used and the metaTree also organizes information in a linear fashion. The DR-tree adopts a ``partitioning'' of space into north, south, east, and inside, which is adopted by all nodes in the tree. The authors, however, do not present an insertion strategy. It appears that their structure is constructed using bulk loading, and it is unclear if the DR-tree can handle changes to the object set without having to be reconstructed. The 2DR-tree proposes a 2-dimensional node structure that fits the data as given and preserves spatial relationships between objects, instead of forcing n-dimensional data to fit a one-dimensional structure. The insertion strategy applies node validity rules to ensure that 1) spatial relationships between objects are maintained, and 2) reconstruction of the entire tree on insertion of a new object is not required. Limitations of the 2DR-tree include a unnecessarily high tree height and a low average space utilization within its nodes, which in turn results in high coverage, overcoverage and overlap of minimum bounding rectangles.

Therefore, we present an improved two-dimensional node and tree structure and strategy for spatial relationship preservation within a node. We also present our insertion and region search algorithms, along with a performance evaluation for both. Our goal is to eliminate the linear nature of existing spatial access methods, and at the same time improve the height, space utilization, overlap and overcoverage of existing methods.

## 3 THE MQR-TREE

In this section, we present our approach to the tree, organizing objects within each node, and the new insertion and search strategies. In addition, we demonstrate some features of the new insertion strategy through a short example. We utilize several terms in our work, which we define here first.

The term *object* can represent any object in two-dimensional space, such as a point, line or polygon of arbitrary shape. An object or subregion of space that contains a subset of objects is approximated using a *minimum bounding rectangle (MBR)*. An MBR defines the minimum two-dimensional rectangular range that an object (or subregion of space) occupies. A *node MBR* is defined for every node in the tree. For a given node, it is the minimum two-dimensional extent that encompasses all MBRs in the node and all of the nodes in its subtrees.

### 3.1 Structure

All nodes in the mqr-tree have the same two-dimensional structure. Fig. 1 depicts the node layout. A node contains 5 locations - northeast (NE), northwest (NW), southwest (SW), southeast (SE) and centre (EQ). Each location contains either:

$$(MBR, obj\_ptr)$$

where *obj_ptr* is a pointer that references an object and *MBR* is the approximation of the object, or:

$$(MBR, node\_ptr)$$

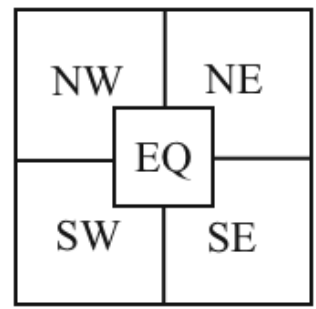

Fig. 1 Node Layout

| Ax = Bx | Ax > Bx | Ay = By | Ay > By | Placement |
|---------|---------|---------|---------|-----------|
| 0 | 0 | 0 | 0 | SW |
| 0 | 0 | 1 | 0 | SW |
| 0 | 0 | 0 | 1 | NW |
| 1 | 0 | 0 | 1 | NW |
| 0 | 1 | 0 | 0 | SE |
| 1 | 0 | 0 | 0 | SE |
| 0 | 1 | 0 | 1 | NE |
| 0 | 1 | 1 | 0 | NE |
| 1 | 0 | 1 | 0 | EQ |

Fig. 2 Orientation of A with respect to B

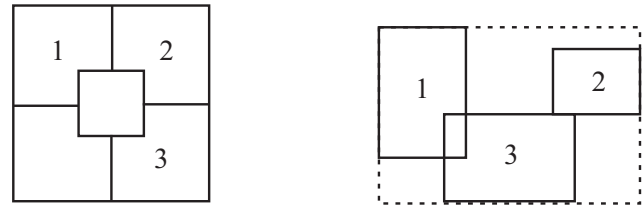

Fig. 3 Node with Objects

where *node_ptr* is a pointer to a subtree and *MBR* is the MBR that encompasses all MBRs in the subtree.

Every node in the mqr-tree must have at least two locations, and not more than five locations, that are referencing either an object or a subtree. It is possible to have a node contains pointers to both objects and subtrees. In light of this, we relax the requirement that the tree must be height-balanced. However, no path in our experiments (see Section 5) is more than 50% longer than the average path length.

### 3.2 Node Organization and Validity

In every node of the mqr-tree, we determine the relative placement of both objects and subregions by using the centroids of their MBRs. We define the origin of each node as its centre location. The objects that are referenced from the centre location have the same centroid as the centroid of the *node MBR* for the node. All other objects and subregions that are referenced from the other locations (NW,SE,SW,NE) are placed with respect to the centroid of the node MBR. Fig. 2 depicts the spatial relationships, where *A* refers to the centroid of a new object, and *B* refers to the centroid of the node MBR. The orientations (NE, SE, SW, NW) include centroids that fall on the axes (E, S, W, N, respectively).

A node is classified as either 'NORMAL' or 'CENTER'. In a NORMAL node, the locations are organized based on the orientations defined above (see Fig. 1). A NORMAL node is valid when:

- The node MBR encloses all the minimum bounding rectangles in the objects or subtrees that the node references, and
- All objects or subtrees pointed to by a location are in the proper quadrant relative to the node centroid.

In a CENTER node, the locations are organized linearly. A CENTER node only references objects whose centroids are the same as the centroid of the node MBR. In addition, a CENTER node is utilized only when more than one object exists with overlapping centroids.

Fig. 3 depicts a NORMAL node containing three objects. Object 1 is located northwest of the centroid of the node MBR (defined by the dashed box on the diagram), while object 2 is located northeast of the centroid of the node MBR. Object 3 is located directly south of the node MBR centroid, therefore it is placed in the southeast quadrant.

Although the mqr-tree node resembles a quadtree node [4], there are some significant differences that must be noted. First, the quadtree defines a recursive partition of space using the points that are inserted. Therefore, the entire space is indexed, including areas that contain no points. The mqr-tree defines an approximation of regions that contain objects. Therefore, the amount of space that is indexed by the mqr-tree is reduced, because only space that contains objects is indexed. Some whitespace (i.e. overcoverage) may be present, but is often significantly less than managed by a partition-based method. Second, the partitions in a quadtree are static after they are created when a point is inserted. In other words, the point that divides the space does not change when other points are inserted. In the mqr-tree, the point that divides the space into quadrants is flexible, and adjusts when objects are inserted. Finally, the mqr-tree accommodates objects whose centroid overlaps the node centroid separately from the node centroid itself, where in the quadtree the data point and the quadrant partitioning point are the same. This allows for the flexibility of the node MBRs (and their centroids) in the mqr-tree.

### 3.3 Insertion Strategy

The insertion strategy works as follows. Beginning at the root node, the node MBR is adjusted to include the new object. Then, the appropriate location, relative to the centroid of the node MBR, is identified for inserting a reference to the new object. If the location is empty, the reference to the object is inserted. Otherwise, the subtree is traversed in the same manner, until either: 1) an appropriate location is found that is empty and the object reference can be inserted, or 2) a leaf node is reached, and no proper location is available for the new object reference. If the object reference cannot be inserted in the proper location of the leaf node, then a new leaf node is created.

In addition, for each node on the insertion path, node validity is maintained by removing and reinserting objects that have changed orientation relative to the centroid of the node MBR during the insertion process. When inserting an object from a node, one of four things will happen to the node MBR:

- The centroid of the node will not change and therefore all objects remain in their proper orientation,
- The node is a CENTER node, and the new object to

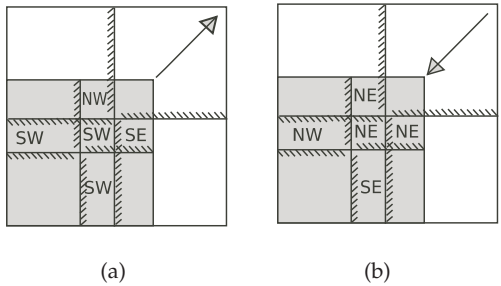

(a)        (b)

Fig. 4 (a) NE Node MBR Expansion (b) SW Node MBR Contraction

be inserted has a different centroid than the existing objects.

- The centroid of the node moves as the region of the node MBR increases in size,
- The centroid of the node moves as the region of the node MBR decreases in size.

In the second case, a CENTER node contains existing points or objects that all have the same centroid, but the new object has a centroid that differs from the other objects or points. Therefore, all existing objects must be moved. In the latter two cases, some objects may have shifted to a different quadrant due to the movement of the centroid of the resulting node MBR. If so, these objects are no longer in their proper relative node location. Any objects in this situation must be located and moved. In all of cases 2-4 above, the objects are moved by re-inserting them beginning at the current level of the tree so that they are placed in a proper relative location.

Fig. 4(a) shows a NE expansion of the node MBR from the original area (shaded) to the new area. The MBR is split into four quadrants using hashed lines. The direction of the hash indicates the quadrant in which the object on that line is included. The regions that are labeled represent the destination location for the objects found within that region.The 'EQ' location has been omitted for clarity. Notice that after the MBR is expanded, partial regions that once belonged to the NW, NE and SE quadrants now belong to the region that makes up the SW quadrant. Any objects within these areas are no longer properly located relative to the new node centre, and would have to be relocated. Fig. 4(b) shows a SW contration in a similar manner. All other expansions and contractions work in a similar manner.

### 3.4 Insertion Implementation Details

Here, we present the implementation details for our insertion strategy. As mentioned earlier, at each level of the insertion path, one or more of the following actions take place when the MBR that represents a new object is inserted into a node:

1) Prepare the new object for insertion into the current node,
   - increase the size of the node MBR of the current

node in order to enclose the new object,
   - determine which quadrant that the reference to the new object will be potentially inserted into,
   - add the MBR reference of the new object to the insertion queue.
2) Locate the MBRs of objects or subtrees whose existing location in the node is no longer valid because the centroid of the node MBR has changed. Add these MBRs to the insertion queue.

Algorithm: insert
Input:
   n: node - node in which to insert newobj
   newobj: object - pointer to new object

Variables:
   objs: queue - queue of objects to be inserted
   orig_mbr: mbr before new object is inserted in n
   item: object to be placed on objs queue

```
=== Begin ===
# if node is empty, insert newobj in center
if number_children(n) == 0
  n->mbr = newobj->mbr
  n->loc[EQ] = newobj

else
  # copy the original node MBR
  orig_mbr = n->mbr

  # merge newobj's MBR into the node's MBR
  merge_mbrs( n->mbr, newobj->mbr )

  # Prep newobj for insertion
  item.quad =
      find_insert_quad( newobj->mbr, n->mbr )
  item.obj = newobj

  # Add newobj to the insertion queue
  enqueue(objs, item)

  # Find other objects that are no longer in a
  # valid quadrant
  # Add them to the insertion queue
  find_shifted_objs(objs, n, orig_mbr)

  # (Re)insert objects in the current node
  insert_queue(n, objs)

return

=== End ===
```

Fig. 5 Main Insertion Strategy

```
Algorithm:  find_shifted_objs

Input:
    q: queue - queue in which to place objects that need
            to be moved
    n: node - node to be updated
    orig_mbr: mbr - MBR of node n before merge

Variables:
    q: quadrant - relative location
    area_diff: int   region: mbr

=== Begin ===
#first, find if any objects have shifted quads
quad = find_insert_quad( n->mbr, orig_mbr )

if quad = EQ
  # nothing to do, MBR may have changed but all
  # objects are already in their proper location
  return

if n->type = CENTER
  # The node is a center node and all objects in this
  # node need to be removed so they can be placed in
  # the appropriate quadrant

  # Find where the objects will be inserted
  quad = find_insert_quad( orig_mbr, n->mbr)

  # remove all objects and place them on the queue
  for each node location 'tmploc'
    remove_and_q_objects(q, quad, tmploc, n->mbr)
  done
  n->type = NORMAL
  return

# Get objects that belong in the EQ location
adjust_region(n->mbr.cx, n->mbr.cx,
            n->mbr.cy, n->mbr.cy)
remove_and_q_objects(q, EQ, n->loc[quad], region)

area_diff = area_change( n->mbr, orig_mbr)
```

Fig. 6 Finding Shifted Objects – Part 1

```
#new node MBR larger than original
if area_diff > 0
  if quad = NE
    # to SE
    adjust_region(n->mbr.cx, orig_mbr->hx,
                orig_mbr->cy, n->mbr.cy -1)
    remove_and_q_objects(q, SE, n->loc[NE], region)

    # to SW, part NE
    adjust_region(orig_mbr->cx +1, n->mbr.cx -1,
      orig_mbr->cy, n->mbr.cy)
    remove_and_q_objects(q, SW, n->loc[NE], region)

    # to SW, part SE
    adjust_region(orig_mbr->cx, n->mbr.cx -1,
      orig_mbr->ly, orig_mbr->cy -1)
    remove_and_q_objects(q, SW, n->loc[SE], region)

    # to SW, part NW
    adjust_region(orig_mbr->lx, orig_mbr->cx,
      orig_mbr->cy +1, n->mbr.cy)
    remove_and_q_objects(q, SW, n->loc[NW], region)

    # to SW part EQ
    remove_and_q_objects(q, SW, n->loc[EQ], orig_mbr)

    # to NW
    adjust_region(orig_mbr->cx +1, n->mbr.cx,
      n->mbr.cy +1, orig_mbr->hy)
    remove_and_q_objects(q, NW, n->loc[NE], region)

    # also have cases for SE, NW and SW,
    # and are handled similarly to NE

  else #new node MBR smaller than original

    # contraction cases for NE, SE, SW
    # and NW handled similar to above

  return

=== End ===
```

Fig. 7 Finding Shifted Objects – Part 2

3) (Re)insert the MBRs on the insertion queue into the current node.

Fig. 5 presents the pseudocode for the main insertion strategy, which depicts an overview of the above sequence of events. Fig. 6 and Fig. 7 present a sketch of the implementation for identifying MBRs of object or subregions that are no longer in the proper quadrant with respect to the centroid of the node MBR. It handles all four cases mentioned above: 1) identifying that no changes have occurred, 2) a CENTER node situation exists, 3) the new node MBR is larger than the previous one before a new object was added, and 4) the new node MBR is smaller than the previous one before a new node was added. Fig. 6 depicts the first two situations. First, to determine if no change to the proper locations of all existing references (plus the new object to be inserted) has occurred, the relative location of the centroid for the new node MBR with respect to the centroid of the original MBR is determined. If the centroids overlap, this means that, although an increase of then node MBR could have taken place, the centroid of the node MBR has not changed and all objects in the node are still in their prop-

Algorithm:  remove_and_q_objects

Input:
    q: queue - insertion queue
    quad: int - destination quad for relocated objects
    loc: node quadrant containing objects for relocation
    region: mbr - region containing objects for relocation

Variables:
    item: queue item
    loctmp: location - location iterator

=== Begin ===
if loc is undefined
    return

# We are always going to quad
item.quad = quad

if loc references a node
    n = loc
    if ( overlaps( n->mbr, region ) )
      for each node location 'tmploc'
      do
        remove_and_q_objects( q, quad, tmploc, region)
      done

      # if any objects are recursively removed here the
      # node MBR must be adjusted, or the node deleted
      adjust_node(loc->parent)

else
# location references an object
    if centroid_within(loc->mbr, region)
        item.obj = loc
        enqueue( q, item )
        loc = undefined

=== End ===

Fig. 8 Queuing Misplaced Objects

Algorithm:  insert_queue

Input:
    n: node -- node to insert queue objects
    q: queue -- queue containing objects

Variables:
    item: item from queue to insert   obj: object
    quad: int -- location index
    ntmp: node -- temporary node

=== Begin ===
while queue is not empty do
    item = dequeue(q)
    quad = item.quad
    obj = item.obj

    if n->type = CENTER
        # see if a location if available
        quad = next_ctr_loc(n)
        if quad is defined
            n->loc[quad] = obj
            sort_ctr(n)
            continue;
        # else, new node is added to the CENTER
        # node list for object

    if n->loc[quad]
        if n->loc[quad] is a node
            insert( n->loc[quad], obj)
            continue
        else
            # Create a new child and insert
            if quad = EQ and num_child(n) = 1
                # convert node to a CENTER node
                n->type = CENTER
                enqueue( q, item )
            else
                ntmp = new_node()
                insert( ntmp, obj )
                insert( ntmp, n->loc[quad] )
                ntmp->parent = n
                n->loc[quad] = ntmp
    else
        # insert at n->loc[quad]
        n->loc[quad]
done

=== End ===

Fig. 9 (Re)-inserting Misplaced Objects

er quadrants.  The search for shifted objects ends here and the reference for the new object can be inserted (via the insert_queue function - see Fig. 9).

   If the centroids do not overlap, then the other cases must be considered.  The next case is determining if the existing node is a CENTER node, and the new object to be inserted has an MBR with a centroid that is not equal to the centroids of the existing MBRs in the node. We chose to handle this situation simply - we identify which quadrant they will be placed in, remove all MBRs from the CENTER node and place them on the insertion queue (via remove_and_q_objects, see Fig. 8), and change the node type to NORMAL.

   Fig. 7 depicts a portion of the last two cases of node MBR expansion and contraction.  Here, we show how a NE expansion (see Fig. 4(a)) is detected and handled. All other cases for both expansion and contraction are handled similarly. The expansion type is determined using the relative location of the new node MBR with respect to the original node MBR (this was calculated earlier). In our example in Fig. 4(a), this expansion is NE, because the centroid for the new node MBR is northeast of the centroid for the original node MBR. After the expansion type is identified, the corresponding subregions that are affected by the node MBR expansion are also identified. Referring back to Fig. 4(a), there are seven subregions that have shifted from one quadrant to another - NE to SE, NE to SW, NE to NW, NW to SW, SE to SW, NE to EQ (not in the figure), and EQ to SW (also, not in the figure). Any MBRs in these subregions will need to be removed and reinserted into the proper quadrant. Therefore, each subregion is identified using the original and new node MBRs, and any MBRs that reside in these regions are removed and added to the insertion queue using remove_and_q_objects (see Fig. 3).

Fig. 8 depicts the pseudocode for the function remove_and_q_objects. This function takes as input the node quadrant that potentially contains MBRs the must be relocated, the subregion that is affected, the destination quadrant for any MBRs that must be moved, and the insertion queue that the affected MBRs must be placed on. Its goal is to remove all MBRs that are accessible from input quadrant and that overlap the given subregion, and place them on the insert queue. Three situations exist: 1) the input quadrant references nothing, in which case the function terminates, 2) the input quadrant contains an MBR of an object, and 3) the input quadrant contains an MBR of a subtree.

If the input quadrant contains an MBR for an object, then a test is performed to see the centroid of the MBR falls within the shifted region. If so, it is removed from the node and added to the insertion queue.

If the input quadrant contains an MBR for a subtree, then two steps must be carried out. The first is to recursively call the function on each quadrant in the subtree, to identify objects that must be removed and reinserted. The second is to adjust the node MBR or delete the node if all MBRs have been removed during this process.

Finally, Fig. 9 depicts the function for inserting the MBRs on the insert queue into the current node (recall that this current node is the node that we started with at the beginning of the sequence of events above). This will attempt to insert the new object, as well as re-insert any removed MBRs, into the quadrant that is specified for each MBR. Again, for each object or MBR being (re)-inserted, several situations exist: 1) the specified quadrant in the node is empty, 2) the node is a CENTER node, 3) the specified quadrant contains MBR for a node, and 4) the specified quadrant contains an MBR for an object.

If the quadrant specified for an object or MBR is empty, then the MBR and corresponding reference can be inserted, and insertion is finished for the current object or MBR. If the node is a CENTER node, then the next available location for the MBR and reference is located and is inserted. If necessary, an additional node is added to form a linked list of CENTER nodes. If the quadrant that is specified for the new object or MBR is referencing a subtree, then the insert function (see Fig. 5) is called on the object or MBR, and the root of the subtree. Finally, if the quadrant contains an MBR for an oboject, a new child node is created, and the insert function is called on both MBRs.

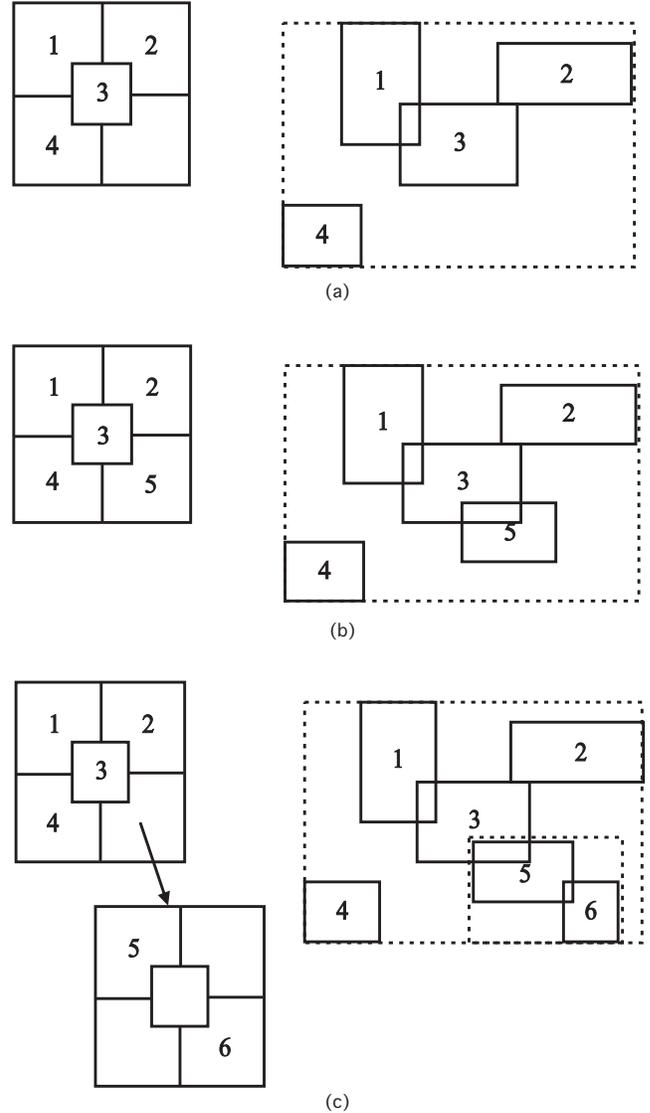

(a)

(b)

(c)

Fig. 10 Insertion Example

### 3.5 Example

Here, we demonstrate a few features of the mqr-tree by inserting some objects. Beginning with the node and objects in Fig. 3, we will insert three more objects into the mqr-tree. First, Object 4 is inserted, which causes the node MBR to increase, Object 3 is no longer be in its proper location because its centroid now overlaps the centroid for the node MBR. Therefore, Object 3 is also reinserted, and is placed in the EQ location. Fig. 10(a) depicts the resulting node.

Next, Object 5 is inserted. The node MBR does not change, so therefore no objects need to be checked to determine their proper location. Fig. 10(b) depicts the result node. Finally, Object 6 is inserted. The southeast quadrant is chosen, and the node MBR does not change. However, object 5 is already located in the same quadrant. Therefore, a new leaf node is created and Objects 5 and 6 are inserted into it. The new node MBR is created and referenced from the southeast quadrant in the parent node. Fig. 10(c) depicts the resulting mqr-tree.

### 3.6 Search

The mqr-tree has the potential for exploring different types of search strategies. For example, the opportunity exists to use a binary partition of nodes for performing a region search. For initial comparison purposes, we chose to implement and use a region search strategy that evaluates the overlap of a search region with the MBRs of all objects or subtrees that are referenced by a node.

## 4 PROPERTIES

In our investigations, we have identified some interesting properties of the mqr-tree index and insertion algorithm:

- For a distinct set of points, any point, and therefore any MBR centroid, has only one possible location in the tree. This leads to a tree that is independent of the insertion order of all objects.

- The centroid of a node will have the same orientation in its parent as does all the objects inclosed by the node MBR.

- The MBR of a location will have less then half of its area outside its quadrant, except for the 'EQ' quadrant. This may lead to a minimizing in overlap.

- With datasets consisting of only points, the overlap of any two MBRs at any level of the tree is zero. There is no area that has the potential to be covered twice.

## 5 EVALUATION

In this section, we present the results of our empirical evaluation of the mqr-tree. Initially, we compared the mqr-tree with the 2DR-tree [13]. We found that the mqr-tree achieved significant improvements in height, space utilization, coverage, overcoverage and overlap over the 2DR-tree. Here, we compare the performance of the mqr-tree insertion and region search algorithms with those from the R-tree [9], which is considered one of the benchmark approaches for spatial indexing. We evaluate the mqr-tree insertion algorith musing rectangles, points, and lines. In particular, line data may generate significantly high amounts of overcoverage because a line is approximated with an MBR. Therefore, it is important to evaluate how the mqr-tree performs in the presence of line data.

### 5.1 Data Sets

We use both synthetic and real datasets for our comparison. With the exception of the road and railroad data (see below), all datasets are randomly generated and contain between 500 and 10,000 objects. Our datasets consist of:

- squares of 10x10 units each, where each dataset assumes a uniform distribution,

- points, where each dataset assumes a uniform distribution,

- squares of 10x10 units each, where each dataset assumes an exponential distribution,

- points, where each dataset assumes an exponential distribution,

- lines of 10 units each, where each dataset contains 50% horizontal and 50% vertical lines,

- lines of 10 units each, where each dataset contains equal percentages of lines of slope 1/2, 1, 2, -1/2, -1 and -2 respectively.

- lines of 10 units each, where each dataset contains equal percentages of lines of slope 1/2, 1, 2, -1/2, -1, -2, 0, horizontal and vertical lines respectively.

- road and railroad data that vary in size from 11,000 to 122,000 line segments. This data is part of the Digital Chart of the World and obtained from [23].

### 5.2 Experiments

For each dataset, we created 100 trees using each insertion algorithm on a random ordering of the dataset. The number of nodes, height, average space utilization in each node, total coverage of all MBRs, total overcoverage (i.e. whitespace) of all MBRs, and the total overlap between all MBRs was calculated for each tree. Because the mqr-tree is not height-balanced, the average path length (i.e. average height) is also recorded. In addition, for each tree created, 20 region searches were performed. Over all 20 searches, the average number of objects that overlapped the search region and the average number of disk accesses required were calculated.

### 5.3 Space Utilization Results

Due to lack of space in our result tables below, we omitted our space utilization calculations. However, over all insertion results, the average space utilization in nodes of the mqr-tree is between 50-55%, while the space utilization for the R-tree was between 70-74%. Although the space utilization of the mqr-tree is 22-23% lower than that of the R-tree, it is still at least half of the number of locations in the node.

### 5.4 Insertion Results on Uniformly Distributed Synthetic Data

Table 1 displays the results for the data sets consisting of uniformly distributed squares. Note that for the R-tree, the values for all parameters are averaged over all 100 runs, since these values vary for each tree. For the mqr-tree, the values are identical for all 100 trees. As mentioned earlier,

#### Table 3 – Uniform Distribution of Objects

| #obj | Index | #node | Height | Coverage | Overcov | Overlap |
|---|---|---|---|---|---|---|
| 500 | mqr-tree | 285 | 8 (5) | 287023.51 | 39892.16 | 21050.87 |
| | r-tree | 196 | 4 | 335178.08 | 59979.97 | 40878.57 |
| 1000 | mqr-tree | 591 | 8 (6) | 626970.98 | 78979.44 | 46605.40 |
| | r-tree | 394 | 4 | 789612.88 | 133872.84 | 102429.08 |
| 5000 | mqr-tree | 2849 | 10 (7) | 3680266.82 | 385573.84 | 245436.34 |
| | r-tree | 1961 | 6 | 5354990.27 | 909009.13 | 783091.64 |
| 10000 | mqr-tree | 5770 | 10 (7) | 7813449.90 | 751530.93 | 487169.05 |
| | r-tree | 3926 | 6 | 12137966.53 | 2203426.85 | 1977476.17 |
| 50000 | mqr-tree | 28807 | 12 (9) | 45170573.09 | 3832563.53 | 2524233.03 |
| | r-tree | 19636 | 7 | 90814453.43 | 16463547.35 | 15451954.96 |
| 100000 | mqr-tree | 57776 | 12 (9) | 95725388.87 | 7717738.72 | 5110715.01 |
| | r-tree | 39255 | 8 | 212272854.03 | 40361580.83 | 38452878.51 |

#### Table 4 – Uniform Distribution of Points

| #pts | Index | #node | Height | Coverage | Overcov | Overlap |
|---|---|---|---|---|---|---|
| 500 | mqr-tree | 281 | 8 (5) | 183202.42 | 52329.78 | 0.00 |
| | r-tree | 191 | 4 | 232603.89 | 70802.17 | 18472.38 |
| 1000 | mqr-tree | 584 | 8 (6) | 412179.74 | 101789.65 | 0.00 |
| | r-tree | 382 | 4 | 575002.63 | 154275.98 | 52486.32 |
| 5000 | mqr-tree | 2880 | 10 (7) | 2608566.97 | 503669.43 | 0.00 |
| | r-tree | 1921 | 6 | 4245530.93 | 1024153.59 | 520484.17 |
| 10000 | mqr-tree | 5758 | 10 (7) | 5683295.65 | 999477.64 | 0.00 |
| | r-tree | 3845 | 6 | 9634100.78 | 2399071.70 | 1399594.05 |
| 50000 | mqr-tree | 28814 | 12 (9) | 34311970.45 | 5017147.64 | 0.00 |
| | r-tree | 19242 | 7 | 80654914.07 | 18486457.58 | 13469309.91 |
| 100000 | mqr-tree | 57737 | 12 (9) | 73778576.93 | 10048704.63 | 0.00 |
| | r-tree | 38526 | 8 | 184978601.88 | 42812880.68 | 32764176.05 |

#### Table 1 – Exponential Distribution of Points

| #pts | Index | #node | Height | Coverage | Overcov | Overlap |
|---|---|---|---|---|---|---|
| 500 | mqr-tree | 325 | 18 (10) | 110974.18 | 52092.32 | 0.00 |
| | r-tree | 186 | 4 | 214523.43 | 64108.84 | 12016.53 |
| 1000 | mqr-tree | 659 | 20 (12) | 218845.42 | 97737.20 | 0.00 |
| | r-tree | 372 | 4 | 472716.17 | 128215.98 | 30478.78 |
| 5000 | mqr-tree | 3370 | 24 (14) | 1660007.95 | 500846.52 | 0.00 |
| | r-tree | 1847 | 6 | 3434431.79 | 800483.10 | 299636.56 |
| 10000 | mqr-tree | 6828 | 26 (15) | 3732725.27 | 999599.36 | 0.00 |
| | r-tree | 3693 | 6 | 7498358.61 | 1828481.67 | 828882.29 |
| 50000 | mqr-tree | 35349 | 30(18) | 50762334.93 | 11287680.72 | 0.00 |
| | r-tree | 18457 | 7 | 122270387.51 | 27474739.89 | 16187059.34 |
| 100000 | mqr-tree | 69693 | 32(19) | 113734895.17 | 22594718.01 | 0.00 |
| | r-tree | 36926 | 8 | 276474613.96 | 61142614.97 | 38547897.18 |

#### Table 2 – Exponential Distribution of Objects

| #obj | Index | #node | Height | Coverage | Overcov | Overlap |
|---|---|---|---|---|---|---|
| 500 | mqr-tree | 325 | 18(10) | 174250.89 | 51192.16 | 18343.37 |
| | r-tree | 194 | 4 | 272040.01 | 64329.05 | 28197.22 |
| 1000 | mqr-tree | 659 | 20(12) | 338443.42 | 95161.33 | 38358.15 |
| | r-tree | 392 | 4 | 606210.34 | 124529.81 | 60858.27 |
| 5000 | mqr-tree | 3370 | 24(14) | 2270464.58 | 488148.33 | 201441.44 |
| | r-tree | 1994 | 6 | 3933855.05 | 797565.45 | 480465.59 |
| 10000 | mqr-tree | 6827 | 26(15) | 4966049.83 | 974169.77 | 402070.90 |
| | r-tree | 4012 | 6 | 8682826.19 | 1761700.06 | 1134572.72 |
| 50000 | mqr-tree | 34711 | 30(18) | 28520871.48 | 4845459.48 | 1977241.75 |
| | r-tree | 20311 | 8 | 61323913.50 | 12127981.69 | 9004376.41 |
| 100000 | mqr-tree | 68910 | 32(19) | 62395597.46 | 9725941.81 | 3942250.98 |
| | r-tree | 40821 | 8 | 137228053.81 | 28266370.27 | 22024751.39 |

the new insertion strategy is independent of the order in which the objects are inserted. The only variation is in how many objects are moved in order to maintain node validity. Also note the two sets of values for height for the mqr-tree. The first value represents the maximum (i.e. worst-case) height, while the second value in parentheses is the average height (i.e. average path length).

Results show that the mqr-tree achieves a significant improvement over the R-tree in many aspects. In particular, there is a 14-55% decrease in coverage, a 33-80% decrease in overcoverage, and a 49-87% decrease in overlap. In all cases, the improvements increase as the number of objects increases. In addition, although the maximum tree height of the mqr-tree is higher than that of the R-tree, the difference in tree height decreases to 33% as the number of objects increases. It must also be noted that the average tree height of the mqr-tree is almost equal to the height of the R-tree. Also, it must be noted that the mqr-tree requires more storage space since it requires 45-50% more nodes than the R-tree. However, we believe that these limitations are a small price to pay for the significant decrease in coverage, overcoverage and overlap, which result in an increase in search performance. Table 2 presents the results for the data sets consisting of uniformly distributed points. The most sig-nificant finding in these results is that *zero overlap is achieved when an index is constructed for points using the mqr-tree insertion strategy.* This is very important because point queries can be executed without having to potentially traverse multiple paths in the tree. In addition, significant reductions in coverage (21-60%) and overcoverage (26-77%) occur. The values for height are similar to those obtained for the object datasets, and therefore we feel these are significantly outweighed by the achievement of zero overlap.

### 5.5 Insertion Results on Exponentially Distributed Synthetic Data

Table 4 presents the results for the data sets containing exponentially distributed objects. Significant decreases in coverage (36-55%), overcoverage (20-70%) and overlap (35-82%) occur in the mqr-tree over the R-tree.

Table 3 presents the results for the data sets consisting of exponentially distributed points. Zero overlap is still achieved, and significant improvements in coverage (48-59%) and overcoverage (18-63%) of the mqr-tree over the R-tree are still achieved. Therefore, it is possible that the mqr-tree can perform a one-path search at most, while multiple search paths may be required for a point search in the R-tree. It should be noted that for the exponential-ly-distributed data sets, the mqr-tree achieves significantly worse tree height - in both the worst and average cases - than the R-tree.

**Table 7 - Road and Railroad Data**

| #obj | Index | #node | Height | Coverage | Overcov | Overlap |
|---|---|---|---|---|---|---|
| **MXrrline** | mqr-tree | 6737 | 12(9) | 1294.95 | 541.89 | 2.20 |
| **10060** | r-tree | 3629 | 6 | 8822.50 | 4074.37 | 3534.69 |
| **CArrline** | mqr-tree | 7755 | 14(9) | 248.10 | 94.13 | 1.01 |
| **11381** | r-tree | 4066 | 6 | 9358.27 | 4442.16 | 4349.06 |
| **CArdline** | mqr-tree | 14116 | 14(9) | 352.01 | 97.20 | 4.19 |
| **21381** | r-tree | 7805 | 7 | 20992.97 | 9561.39 | 9468.41 |
| **CDrrline** | mqr-tree | 23108 | 14(10) | 4324.47 | 1561.67 | 18.18 |
| **35074** | r-tree | 12564 | 7 | 35090.63 | 15508.73 | 13965.37 |
| **MXrdline** | mqr-tree | 58851 | 14(10) | 2038.99 | 553.05 | 16.87 |
| **92392** | r-tree | 32872 | 8 | 93816.25 | 41718.96 | 41182.92 |
| **CDrdline** | mqr-tree | 76998 | 16(11) | 9500.39 | 2925.65 | 59.95 |
| **121416** | r-tree | 43438 | 8 | 133233.80 | 55883.37 | 53017.86 |

**Table 5 - Sloped Lines**

| #obj | Index | #node | Height | Coverage | Overcov | Overlap |
|---|---|---|---|---|---|---|
| **500** | mqr-tree | 284 | 8(5) | 271315.61 | 50635.87 | 15181.68 |
| | r-tree | 195 | 4 | 319704.04 | 69274.30 | 34247.39 |
| **1000** | mqr-tree | 581 | 8(6) | 586703.65 | 100469.16 | 34995.16 |
| | r-tree | 392 | 4 | 763931.88 | 152911.51 | 88912.03 |
| **5000** | mqr-tree | 2907 | 10(7) | 3506656.31 | 492314.23 | 188213.63 |
| | r-tree | 1963 | 6 | 5144578.40 | 1015634.70 | 721374.75 |
| **10000** | mqr-tree | 5760 | 10(7) | 7445892.16 | 978976.24 | 386760.43 |
| | r-tree | 3920 | 6 | 11572362.19 | 2395879.76 | 1826034.50 |
| **50000** | mqr-tree | 28909 | 12(9) | 43421521.54 | 4935490.52 | 2017537.19 |
| | r-tree | 19608 | 7 | 87460073.45 | 17320731.04 | 14544767.33 |
| **100000** | mqr-tree | 57942 | 12(9) | 91993440.89 | 9851272.71 | 4044739.58 |
| | r-tree | 39216 | 8 | 206787354.04 | 42502264.26 | 37005936.91 |

**Table 8 - Horizontal and Vertical Lines**

| #obj | Index | #node | Height | Coverage | Overcov | Overlap |
|---|---|---|---|---|---|---|
| **500** | mqr-tree | 293 | 8(5) | 251619.56 | 67535.15 | 10782.32 |
| | r-tree | 195 | 4 | 291630.53 | 83903.64 | 27150.81 |
| **1000** | mqr-tree | 587 | 8(6) | 544531.26 | 132311.38 | 24121.85 |
| | r-tree | 394 | 4 | 696390.31 | 185733.09 | 77543.57 |
| **5000** | mqr-tree | 2914 | 10(7) | 3294526.16 | 655888.20 | 137649.08 |
| | r-tree | 1961 | 6 | 5042184.13 | 1241062.37 | 722823.28 |
| **10000** | mqr-tree | 5810 | 10(7) | 7001261.27 | 1292190.54 | 272790.93 |
| | r-tree | 3929 | 6 | 10965166.53 | 2715467.58 | 1696067.96 |
| **50000** | mqr-tree | 28890 | 12(9) | 41092930.60 | 6506622.52 | 1444385.16 |
| | r-tree | 19621 | 7 | 86853329.18 | 19697559.43 | 14635322.12 |
| **100000** | mqr-tree | 58068 | 14(9) | 87417416.02 | 13027973.93 | 2915851.38 |
| | r-tree | 39224 | 8 | 203407727.08 | 47028991.84 | 36916869.57 |

**Table 6 - Sloped, Horizontal and Vertical Lines**

| #obj | Index | #node | Height | Coverage | Overcov | Overlap |
|---|---|---|---|---|---|---|
| **500** | mqr-tree | 289 | 8(5) | 277018.50 | 56623.55 | 16207.52 |
| | r-tree | 194 | 4 | 324332.16 | 76640.31 | 36806.05 |
| **1000** | mqr-tree | 572 | 8(6) | 580037.86 | 108356.88 | 33389.45 |
| | r-tree | 392 | 4 | 734513.29 | 152146.81 | 78014.48 |
| **5000** | mqr-tree | 2903 | 10(7) | 3486989.74 | 537944.78 | 184403.61 |
| | r-tree | 1963 | 6 | 5130453.08 | 1062417.04 | 716194.67 |
| **10000** | mqr-tree | 5818 | 10(7) | 7371817.94 | 1051413.83 | 363377.44 |
| | r-tree | 3927 | 6 | 11439524.18 | 2458147.81 | 1786487.04 |
| **50000** | mqr-tree | 28979 | 12(9) | 42915843.71 | 5285883.19 | 1887563.59 |
| | r-tree | 19624 | 7 | 87958140.65 | 18005334.35 | 14704704.00 |
| **100000** | mqr-tree | 58083 | 12(9) | 91056946.11 | 10614180.99 | 3835949.08 |
| | r-tree | 39229 | 8 | 205952536.38 | 43688852.06 | 37122966.54 |

### 5.6 Insertion Results on Road and Railroad Data

Table 5 presents the results for the road and railroad data. Here, we also achieve almost no overlap in the mqr-tree. The results show that a reduction in overlap of almost 100% is achieved. In addition, we also achieve over 85% in reduction in both coverage and overcoverage. The mqr-tree has a higher tree height in the worst case. However, given that the overlap and overcoverage of the mqr-tree are significantly low, it is expected that more efficient searching will be achieved despite the higher tree height.

### 5.7 Insertion Results on Synthetic Line Data

The results from the road and railroad data sets were surprising. We conclude that the above results occurred because: 1) the line segments were very small, and 2) the line segments were sequential in nature (for example, a road is made of several line segments that are connected end-to-end). Therefore, we conducted further experiments with randomly-generated line sets, where the lines are much longer than those in the road and railroad data sets.

Table 6 presents the results for the horizontal and ver-tical line sets. This is expected to produce the best results since at the leaf level, the overcoverage of minimum bounding rectangles will be zero and the overlap will be very low (effectively, the only overlap are the intersection points between two lines). We find the most significant result to be in the improvement in overlap. The mqr-tree achieves lower overlap in all cases. Although the improvement amounts are not as high as with the road and railroad data, they are still significant, especially in the data sets with the higher number of line segments. Overall, we find in the smaller sets an improvement of approximately 45-50% lower overlap over the R-tree, while in the larger sets the improvement is as high as 92%. We also find the same trends for coverage and overcoverage, with improvements that increase from 3% to 58% for coverage and from 9% to 73% for overcoverage. The height, although still high in the mqr-tree, is comparable to those obtained in the initial road and railroad data tests.

Table 7 depicts the results for the worst-case scenario, where the MBRs for all lines contain a significant amount of overcoverage. However, we find that for the most part the performance improvements of the mqr-tree over the R-tree are as significant as those found in the other evaluations. The only improvement that is not as significant

#### Table 9 - Uniform Objects

| #obj | Index | #found | #diskhits |
|---|---|---|---|
| 500 | mqr-tree | 3.6 | 7.85 |
| | r-tree | 3.6 | 8.089 |
| 1000 | mqr-tree | 4.1 | 8.85 |
| | r-tree | 4.1 | 9.779 |
| 5000 | mqr-tree | 4.05 | 10.05 |
| | r-tree | 4.05 | 13.1845 |
| 10000 | mqr-tree | 4.25 | 11.35 |
| | r-tree | 4.25 | 14.9075 |
| 50000 | mqr-tree | 4.55 | 13 |
| | r-tree | 4.55 | 21.615 |
| 100000 | mqr-tree | 3.65 | 13 |
| | r-tree | 3.65 | 24.764 |

#### Table 10 - Uniform Points

| #pts | Index | #found | #diskhits |
|---|---|---|---|
| 500 | mqr-tree | 0.40 | 4.65 |
| | r-tree | 0.40 | 6.02 |
| 1000 | mqr-tree | 0.85 | 5.75 |
| | r-tree | 0.85 | 7.91 |
| 5000 | mqr-tree | 0.75 | 7.10 |
| | r-tree | 0.75 | 11.17 |
| 10000 | mqr-tree | 1.10 | 7.75 |
| | r-tree | 1.10 | 13.31 |
| 50000 | mqr-tree | 1.30 | 9.25 |
| | r-tree | 1.30 | 20.96 |
| 100000 | mqr-tree | 1.25 | 10.25 |
| | r-tree | 1.25 | 24.00 |

#### Table 11 - Exponential Objects

| #obj | Index | #found | #diskhits |
|---|---|---|---|
| 500 | mqr-tree | 273.00 | 180.85 |
| | r-tree | 273.00 | 113.95 |
| 1000 | mqr-tree | 539.30 | 364.05 |
| | r-tree | 539.30 | 224.38 |
| 5000 | mqr-tree | 2595.50 | 1780.65 |
| | r-tree | 2595.50 | 1074.92 |
| 10000 | mqr-tree | 5122.05 | 3529.81 |
| | r-tree | 5122.05 | 2121.67 |
| 50000 | mqr-tree | 24926.10 | 17543.35 |
| | r-tree | 24926.10 | 10418.19 |
| 100000 | mqr-tree | 49275.10 | 34282.90 |
| | r-tree | 49275.10 | 20697.80 |

#### Table 12 - Exponential Points

| #pts | Index | #found | #diskhits |
|---|---|---|---|
| 500 | mqr-tree | 90.50 | 68.85 |
| | r-tree | 90.50 | 71.16 |
| 1000 | mqr-tree | 172.70 | 132.50 |
| | r-tree | 172.70 | 133.83 |
| 5000 | mqr-tree | 755.70 | 534.25 |
| | r-tree | 755.70 | 577.42 |
| 10000 | mqr-tree | 1440.80 | 1014.75 |
| | r-tree | 1440.80 | 1101.89 |
| 50000 | mqr-tree | 6451.80 | 4629.30 |
| | r-tree | 6451.80 | 4958.53 |
| 100000 | mqr-tree | 12333.05 | 8888.35 |
| | r-tree | 12333.05 | 9628.49 |

is in the overlap decrease for the smallest test set. However, the mqr-tree still achieves lower overlap in this case.

Table 8 presents the results of our average case, where there is a mix of lines that result in significant overcoverage and lines that have no overcoverage. We find results that are very similar to those for the horizontal and vertical line sets. We find that the mqr-tree achieves an improvement in overlap that ranges from 43% for the smaller data sets to 90% for the largest one. Similarly, we find improvements in coverage that fall between 3% and 57%, and for overcoverage that fall between 14% and 77%. This is very reassuring because it appears that diagonal lines (which in this case, make up 3/4ths of each data set) do not significantly affect the performance criteria.

### 5.8 Search Results

We compare the mqr-tree region search with the R-tree region search to determine if the significant reduction in overlap and overcoverage results in a significant reduction in the number of disk accesses required for performing the same region search.

Tables 9, 10, 11 and 12 present the results for performing a region search on mqr-trees that were built with uniformly-distributed objects, uniformly-distributed points, exponentially-distributed objects and exponentially-distributed points respectively. For the most part, the mqrtree achieves a lower average number of disk accesses over the R-tree when performing a region search. In particular, when the number of uniformly-distributed objects increases, the amount of improvement significantly increases to over 50%. The exception to the improvement of the mqr-tree on region searching occurs in trees containing exponentially-distributed objects. Here, the R-tree achieves the more significant improvements in the number of disk accesses over the mqr-tree.

## 7 Conclusion

We propose the mqr-tree, a two-dimensional index structure that utilizes more efficient organizational structure than other existing strategies. In addition, it utilizes an insertion algorithm that achieves lower overlap and overcoverage, which in turn achieves improved search performance. We show through experimental evaluation that the mqr-tree outperforms a benchmark indexing strategy, and achieves no or little overlap. In particular, zero overlap is achieved when the mqr-tree is used to index point data, which is not achieved by the R-tree.

Currently, the mqr-tree is limited to two dimensions. Future work includes the following. The first is to investigate the extension of the node structure to multiple dimensions and determine how this affects the search performance of the mqr-tree. The second is to increase the number of locations in a 2-dimensional node. The third, related to the second, is to explore a bottom-up strategy for height balancing. The fourth is to create a bottom-up tree-construction strategy to handle multiple insertions at once. The final improvement is a paging strategy that groups nodes based on a high probability that they are retrieved for the same queries. —these should be referenced in the body of the paper.

### Acknowledgment

This work was supported in part by a grant from the University of Lethbridge Research Fund.

**Marc Moreau** obtained his B.Sc. degree from the University of Lethbridge in 2011. He is currently a system administrator for the City of Calgary, Alberta, Canada. Marc has previously published four conference papers in the areas of spatial indexing and image processing.

**Wendy Osborn** obtained her B.C.S. (Hons.) and M.Sc. degrees from the University of Windsor, and her Ph.D. degree from the University of Calgary. She is currently an Assistant Professor of Computer Science at the University of Lethbridge. She has over 20 papers published in the areas of distributed query optimization, spatial indexing, spatial query processing, digital libraries and context-aware mobile systems.